\documentclass[fleqn,usenatbib]{mnras}

\usepackage{newtxtext,newtxmath}
\usepackage{graphicx}
\usepackage{amsmath}
\usepackage{booktabs}
\usepackage{siunitx}
\usepackage{xcolor}

\newcommand{\rateunit}{\times10^{-3}\,\mathrm{deg\,cycle^{-1}}}
\newcommand{\BJD}{\mathrm{BJD}_{\mathrm{TDB}}}

\title[Stellar tidal systematics in CBP searches]
{Stellar tidal systematics in apsidal-motion searches for circumbinary planets: CH Ind and SW CMa}

\author[Q. Jiang]{
Qunfeng Jiang$^{1}$\thanks{E-mail: qfjiang01@gmail.com; ORCID: 0009-0008-0600-3743}\\
$^{1}$Independent Researcher, Shanghai, China
}

\date{Accepted XXX. Received YYY; in original form ZZZ}
\pubyear{2026}

\begin{document}
\label{firstpage}
\pagerange{\pageref{firstpage}--\pageref{lastpage}}
\maketitle

\begin{abstract}
\citet{Thornton2026} reported 27 non-transiting circumbinary-planet candidates from excess apsidal motion in TESS eclipsing-binary timings. The classical stellar contribution is sensitive to target-specific radii, apsidal constants $k_2$, and rotation. We re-evaluate two evolved, double-lined eclipsing binaries in that sample---the pulsating system CH~Ind and the Am system SW~CMa---whose component masses and radii have been measured in dedicated binary studies, allowing the stellar contribution to be calculated directly. For CH~Ind, six independent TESS sector nodes give $\dot\omega_{\rm obs}=1.82^{+0.99}_{-0.94}\,\rateunit$, while a coeval fit to the measured binary components predicts $\dot\omega_{\rm stars}=2.05^{+0.43}_{-0.34}\,\rateunit$. For SW~CMa, its measured dimensions raise the classical term by a factor of about 16; the published target-specific prediction, $0.670\pm0.020\,\rateunit$, agrees with the observed $0.690\pm0.050\,\rateunit$. Neither system shows a significant excess prograde component. The comparison shows that candidates from a bulk search require target-specific binary models before a third-body interpretation is assigned.
\end{abstract}


\section{Introduction}

Transits have established the existence of circumbinary planets (CBPs), but select predominantly compact, nearly coplanar architectures \citep{Doyle2011,MartinTriaud2015,Kostov2021}. Radial velocities and eclipse-timing variations offer complementary access to non-transiting companions \citep{Standing2023,Goldberg2023}. A third body also drives secular apsidal motion of an eccentric inner binary \citep{Harrington1968}: as its argument of periastron $\omega$ changes, primary and secondary eclipses drift in opposite directions. The Transiting Exoplanet Survey Satellite (TESS; \citealt{Ricker2015}) therefore makes apsidal motion a promising all-sky dynamical search channel.

The measured rate is not a planet observable alone. General relativity (GR), tidal distortion, and rotational flattening all advance the stellar apsides \citep{Sterne1939,Gimenez1985,ClaretGimenez1993}. The classical terms scale as $(R/a)^5$ and depend linearly on the internal-structure constant $k_2$ and quadratically on spin in the rotational contribution. Eclipsing binaries have consequently long served as precision tests of stellar interiors, but only when their component masses, radii, temperatures, and rotation are measured \citep{TorresReview2010,Claret2021}.

\citet{Thornton2026} searched 1590 Gaia eclipsing binaries and reported 27 planetary-mass candidates after subtracting bulk GR and classical estimates. Their efficient discovery-stage calculation combined catalogue temperatures \citep{Stassun2019} with main-sequence relations \citep{Eker2018}, adopted $k_2=0.01$, and assumed synchronous rotation. We test the validation stage for two candidates with unusually informative prior work: CH~Ind has a recent double-lined eclipsing-binary and pulsation analysis \citep{Liakos2025}, while SW~CMa has precise component dimensions \citep{Torres2012} and an already published target-specific apsidal test \citep{Claret2021}. A radius-sensitivity diagnostic for the full candidate list places these two cases in the context of the wider search.

\section{Method}
\label{sec:method}

We downloaded TESS light curves from the Quick-Look Pipeline (QLP; \citealt{Huang2020}) high-level science products and measured every usable primary and secondary eclipse. All new times are on $\BJD$; primary denotes the deeper eclipse. For CH~Ind the preferred result uses the \texttt{SAP\_FLUX} column, empirical eclipse templates, local linear continua, and subtraction of its two strongest independent pulsation frequencies. Repeated eclipses are summarized into one primary and one secondary timing node per sector. Defining
\begin{equation}
 \begin{split}
 D(t)&\equiv\frac{(O-C)_{\rm pri}-(O-C)_{\rm sec}}{2}
 \simeq-\frac{P}{\pi}e\cos\omega(t),\\
 \dot\omega&=\frac{180}{e\sin\omega}\frac{{\rm d}D}{{\rm d}t},
 \end{split}
 \label{eq:timing}
\end{equation}
the last expression is in degrees per binary cycle when $D$ and $t$ use the same time unit. We fit the six $D$ nodes with a line and a shared non-negative sector jitter. Leave-one-sector-out, QLP, alternative detrending, 10-frequency pre-whitening, and a joint first-order fit of all eclipse times provide robustness tests. The complete template construction, priors, and uncertainties are given in the Supporting Information (SI). The long-term slope is therefore constrained by six observing epochs, not by 45 independent eclipse times.

We decompose
\begin{equation}
 \dot\omega_{\rm obs}=\dot\omega_{\rm GR}+\dot\omega_{\rm rot}
 +\dot\omega_{\rm tide}+\dot\omega_{\rm extra}.
 \label{eq:budget}
\end{equation}
With masses in $\mathrm{M}_{\odot}$ and the anomalistic period $P$ in days,
\begin{equation}
 \dot\omega_{\rm GR}=5.4471273\times10^{-4}
 \frac{(M_1+M_2)^{2/3}}{P^{2/3}(1-e^2)}\quad\mathrm{deg\,cycle^{-1}}.
 \label{eq:gr}
\end{equation}
For aligned stellar spins, the classical rate is \citep{Sterne1939,Baroch2021}
\begin{align}
 \dot\omega_{\rm CL}&=360\sum_{i=1}^{2} k_{2,i}
 (c_i^{\rm rot}+c_i^{\rm tide}),\\
 c_i^{\rm rot}&=r_i^5(1-e^2)^{-2}
 \left(1+\frac{M_j}{M_i}\right)\left(\frac{\Omega_i}{n}\right)^2,\\
 c_i^{\rm tide}&=15r_i^5\frac{M_j}{M_i}
 \frac{1+1.5e^2+0.125e^4}{(1-e^2)^5},
 \label{eq:classical}
\end{align}
where $r_i=R_i/a$, $j\ne i$, and $n=2\pi/P$. Here $k_2$ is the apsidal-motion constant tabulated by stellar models, not the Love-number convention that is larger by a factor of two. We propagate masses, radii, $e$, spin, and $k_2$ jointly. Reported excess limits condition the posterior on $\dot\omega_{\rm extra}\geq0$. They therefore apply to the prograde, nearly coplanar configurations motivating the candidate interpretation, not to inclined, retrograde, or Kozai--Lidov regimes \citep{Kozai1962,Lidov1962}.

\begin{figure*}
 \includegraphics[width=\textwidth]{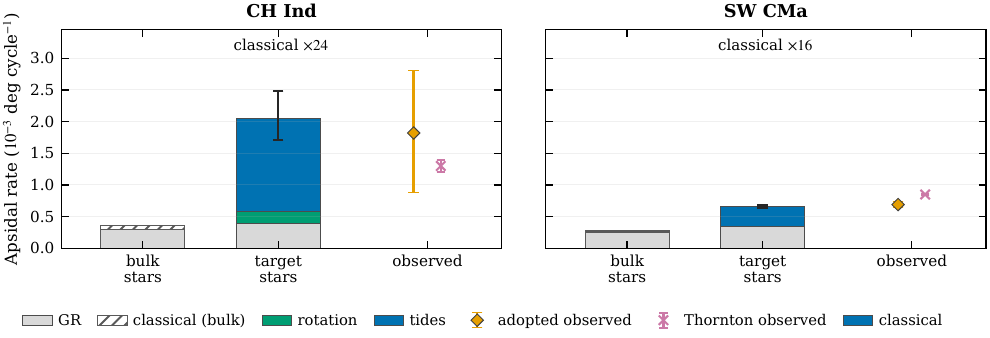}
 \caption{Published bulk estimates from Table~1 of \citet{Thornton2026}, target-specific stellar predictions, and the adopted observations. The CH~Ind target bar separates rotation and tides because separate $k_{2,i}$ values are inferred. For SW~CMa only the total classical term is shown because the published $\bar{k}_2$ is an effective weighted mean. Orange diamonds are the rates adopted here and pink crosses are the rates reported by \citet{Thornton2026}. The classical contributions increase by factors of 24 and 16, respectively.}
 \par\smallskip\noindent\textit{Alt text:} Comparison of bulk, target-specific, and observed apsidal-motion rates for CH~Ind and SW~CMa.
 \label{fig:budget}
\end{figure*}

\section{CH Ind}
\label{sec:chind}

CH~Ind is a $P=5.95257$-d eccentric binary. The detailed binary analysis of \citet{Liakos2025} gives $(M_1,M_2)=(1.82\pm0.01,1.85\pm0.02)\,\mathrm{M}_{\odot}$, $(R_1,R_2)=(3.05\pm0.09,2.79\pm0.09)\,\mathrm{R}_{\odot}$, and $e=0.058\pm0.005$. Both stars are close to the terminal-age main sequence, rather than ordinary unevolved stars. The same work finds 46 pulsation frequencies, with a $\gamma$~Dor primary and a $\delta$~Sct secondary. These facts directly affect both sides of equation~\eqref{eq:budget}: enlarged radii amplify the stellar quadrupoles, while pulsations can perturb individual eclipse times.

\begin{figure*}
 \includegraphics[width=\textwidth]{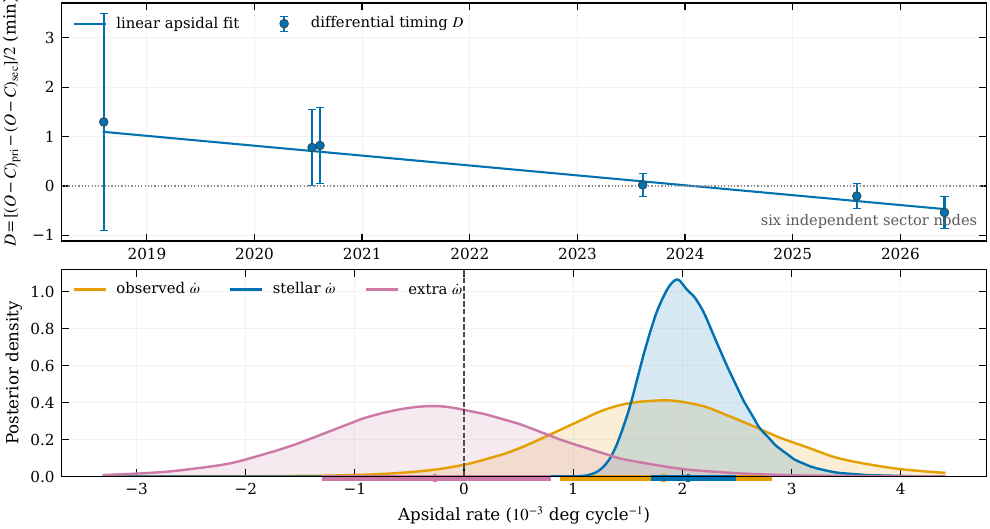}
 \caption{CH~Ind timing and inference. Top: the six differential nodes $D$ from equation~\eqref{eq:timing}; subtracting primary and secondary O--C values removes their common ephemeris or LTTE trend and isolates the first-order apsidal signal. Bottom: Monte Carlo posteriors for the displayed 0.08-dex common $\log k_2$ systematic; horizontal segments show central 68 per cent intervals.}
 \par\smallskip\noindent\textit{Alt text:} CH~Ind differential eclipse timings and posterior distributions of the observed, stellar, and excess apsidal-motion rates.
 \label{fig:chind}
\end{figure*}

We obtain 45 eclipse times in TESS sectors 1, 27, 28, 68, 95, and 104. Their antisymmetric sector nodes show the expected opposing drift after the common ephemeris trend is removed (Fig.~\ref{fig:chind}). The preferred sector-jitter posterior is
\begin{equation}
 \dot\omega_{\rm obs}=1.82^{+0.99}_{-0.94}\,\rateunit.
 \label{eq:chobs}
\end{equation}
Leaving out each sector gives central values from 1.62 to $1.96\,\rateunit$, all with uncertainties of about $0.8$--$0.9\,\rateunit$. Raw and two-frequency-cleaned \texttt{SAP\_FLUX} results differ by $0.02\,\rateunit$; a 10-frequency test gives $1.95\,\rateunit$. The QLP detrended columns, \texttt{KSPSAP\_FLUX} where available and \texttt{DET\_FLUX} otherwise, give 2.09--$2.18\,\rateunit$. We add their method-to-method scatter in quadrature rather than select the smallest formal error. A direct joint first-order timing model gives $1.98^{+0.81}_{-0.72}\,\rateunit$, consistent with the differential result. The more precise $1.30\pm0.10\,\rateunit$ reported by \citet{Thornton2026} cannot be exactly reconstructed because their selected individual CH~Ind timings are not published; the principal difference is that our uncertainty is controlled by six sector nodes and method scatter, not 45 independent events.

We infer $k_{2,1}$ and $k_{2,2}$ from the MESA evolutionary grids of \citet{Claret2023}, fitting both stars at a shared age and metallicity while varying their measured masses and fitting both radii and temperatures. The posterior medians are $k_{2,1}=0.00272$ and $k_{2,2}=0.00243$. We adopt a common 0.08-dex $\log k_2$ systematic for the displayed result and test 0.05--0.20 dex; this is a sensitivity assumption, not a measured model error. Specifically, one shift $\Delta\log_{10}k_2\sim\mathcal{N}(0,\sigma_{\rm sys}^{2})$ is applied to both stars in each Monte Carlo draw. With $\Omega_i/n=1.0\pm0.1$, the component budget is GR $0.396$, rotation $0.193$, and tides $1.465$, yielding
\begin{equation}
 \dot\omega_{\rm stars}=2.05^{+0.43}_{-0.34}\,\rateunit.
 \label{eq:chstars}
\end{equation}
A shared age and composition are the physically appropriate constraints for a binary, but the best-fitting model does not reproduce every measurement exactly: $\chi^2=8.79$ for 3 degrees of freedom ($p=0.032$). In particular, the nominally more massive star is observed to be the smaller one, whereas the models generally predict the opposite ordering at fixed age and composition. We therefore treat the inferred $k_2$ values as model-dependent estimates rather than precision measurements. Matching the stars separately, or changing the treatment of their mass errors, leaves the stellar and observed rates consistent (SI).

Combining equations~\eqref{eq:chobs} and \eqref{eq:chstars} gives
$\dot\omega_{\rm extra}=-0.26^{+1.06}_{-1.04}\,\rateunit$ (Fig.~\ref{fig:chind}). The excess is not detected. Conditioning on a non-negative prograde term and taking the conservative envelope over 0.05--0.20 dex gives $\dot\omega_{\rm extra}<2.28\,\rateunit$ at 95 per cent credibility. We do not add the seven \citet{Liakos2025} minima because they are Sector~68 TESS measurements already represented here; the only other catalogued historical entry lacks a primary/secondary type and uncertainty.

The target-specific stellar prediction is consistent with the sector-level TESS measurement. At the present timing precision, CH~Ind therefore shows no evidence for an additional apsidal contribution.

\citet{Thornton2026} also reported a tentative 2.0-min light-travel-time effect (LTTE) with a period of 10\,056~d. With the measured CH~Ind mass, this corresponds to a minimum companion mass of $0.063\,\mathrm{M}_{\odot}$. In a circular, nearly coplanar orbit, such a companion would contribute only $1.6\times10^{-6}$~deg cycle$^{-1}$ and could not account for the nominal apsidal excess. Because TESS spans less than one third of the proposed outer period, the available data do not constrain a new LTTE solution; the calculation and its limitations are given in the SI.

\section{SW CMa}
\label{sec:swcma}

SW~CMa is an evolved, eccentric, double-lined Am binary with $P=10.091988$~d. \citet{Torres2012} measured $(M_1,M_2)=(2.239\pm0.014,2.104\pm0.018)\,\mathrm{M}_{\odot}$, $(R_1,R_2)=(3.014\pm0.020,2.495\pm0.042)\,\mathrm{R}_{\odot}$, and projected rotations of $24.0\pm1.5$ and $10.0\pm1.0\,\mathrm{km\,s^{-1}}$. The two rotation rates correspond to $\Omega/n=1.59$ and 0.80, so neither a common synchronous nor a common pseudo-synchronous prescription is adequate \citep{Hut1981}.

\citet{Torres2012} obtained the theoretical, coefficient-weighted value $\log\bar{k}_{2,{\rm theo}}=-2.582\pm0.050$; this was a stellar-model prediction, not a value inferred from the observed apsidal rate. \citet{Claret2021} later obtained $-2.581\pm0.015$. Because neither study published two individual constants, we apply an effective $\bar{k}_2$ only to the summed classical coefficient. At the \citet{Torres2012} central value the result is GR $0.345\,\rateunit$ plus a total classical contribution of $0.319\,\rateunit$, or $0.664\,\rateunit$ in total. Any rotation--tide split made with a common $\bar{k}_2$ is diagnostic, not a measurement of the two pieces. The total reproduces the target-specific $0.670\pm0.020\,\rateunit$ prediction of \citet{Claret2021} to 0.87 per cent. By contrast, \citet{Thornton2026} used GR $0.253\,\rateunit$ and classical $0.020\,\rateunit$; the latter is underestimated by a factor of 15.97 (Fig.~\ref{fig:budget}).

For SW~CMa, the relevant apsidal measurement is already available. \citet{Clausen2008} found $0.00067\pm0.00021$ deg cycle$^{-1}$ from archival minima, and \citet{Claret2021} combined reliable historical timings with TESS to obtain
\begin{equation}
 \dot\omega_{\rm obs}=0.690\pm0.050\,\rateunit,
\end{equation}
consistent with the stellar prediction. Our 18 TESS eclipse times add a later sector, but only four independent sector nodes. A simplified fit to the precise historical subset and those nodes gives $0.744\,\rateunit$. Given its reduced $\chi^2=4.42$ and the undocumented time standard of the early HJD minima, we use this fit only as a consistency check and retain the rate of \citet{Claret2021} for the apsidal budget (SI). Combining the published observed rate with a Monte Carlo centred on the \citet{Claret2021} theoretical $\bar{k}_2$ and a conservative 0.05-dex common systematic gives $\dot\omega_{\rm extra}=0.020^{+0.063}_{-0.066}\,\rateunit$ and a non-negative prograde 95 per cent upper limit of $0.14\,\rateunit$. The target-specific stellar calculation therefore accounts for the observed rate, and the excess reported by \citet{Thornton2026} is not significant.

\section{Implications for candidate validation}
\label{sec:implications}

Neither proposed excess remains significant after target-specific stellar subtraction. In both cases, the change comes from the stellar budget rather than the timing signal itself. \citet{Thornton2026} used catalogue temperatures and unevolved main-sequence relations for systems whose measured component radii reach $2.5$--$3.0\,\mathrm{R}_{\odot}$. Because the classical coefficients contain $(R/a)^5$, radius factors of 1.5, 1.8, and 2.0 multiply the result by 7.6, 18.9, and 32. Their fixed $k_2=0.01$ is about four times the target-specific CH~Ind values and partly offsets the radius error, but not enough: the resulting classical terms remain too small by factors of 24 and 16. The GR estimates are also about 27 per cent below the target-specific values for both systems; with $P$ and $e$ fixed, the $M_{\rm tot}^{2/3}$ scaling corresponds to total masses near 63 per cent of the measured values.

The statement that the search is largely insensitive to stellar radius \citep{Thornton2026} is appropriate for recognizing an apsidal timing drift, but not for deciding whether the residual is planetary: that decision requires subtracting the explicitly radius-dependent stellar term. Bulk relations are therefore useful for selecting objects for follow-up, but are not a final false-positive test. CH~Ind requires new sector-aware timing and stellar calculations. For SW~CMa, published target-specific analyses already found that GR, tidal distortion, and rotational flattening reproduce the observed apsidal rate without an additional third-body term. Our independent calculation from its measured masses, radii, rotations, and theoretical $\bar{k}_2$ reproduces the published GR-plus-classical prediction to within 0.87 per cent. Modern validation should start from double-lined binaries with precise dimensions \citep{TorresReview2010,Claret2021}; rotation can also affect the quadrupole and inferred internal constant \citep{Claret2024}.

\begin{figure*}
 \includegraphics[width=\textwidth]{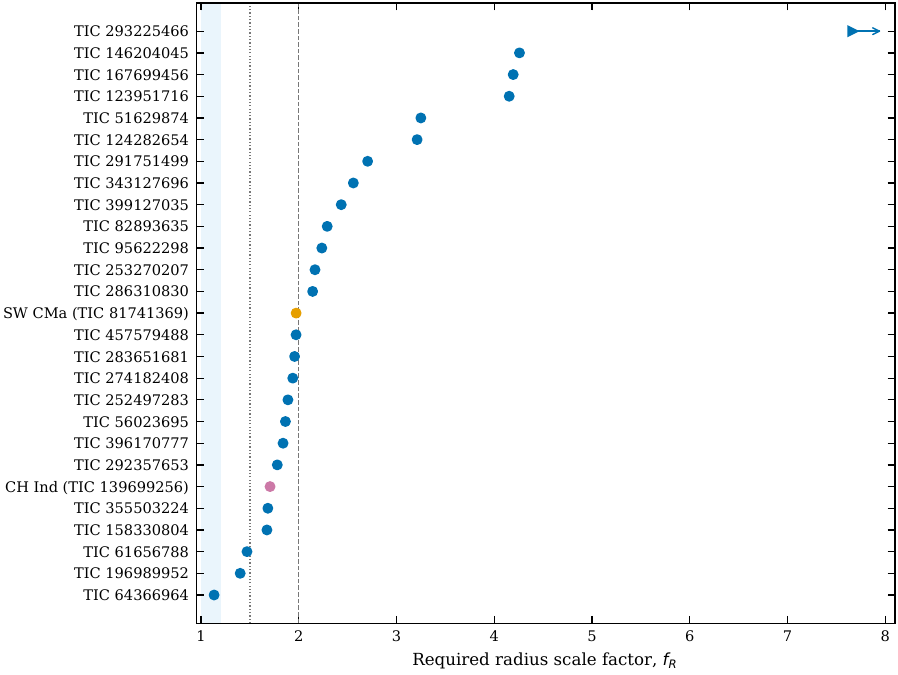}
 \caption{Radius-sensitivity screen for the 27 final candidates of \citet{Thornton2026}, sorted by $f_R$. CH~Ind and SW~CMa are highlighted; the arrow marks the object whose bulk classical term was rounded to zero. The statistic holds $k_2$, rotation, and component weights fixed and is a prioritization aid, not a reclassification.}
 \par\smallskip\noindent\textit{Alt text:} Required stellar-radius scale factors for the 27 candidates, highlighting CH~Ind and SW~CMa.
 \label{fig:fr}
 \label{lastpage}
\end{figure*}

We quantify the radius sensitivity of all 27 candidates with
\begin{equation}
 f_R=\left[\frac{\dot\omega_{\rm obs}-\dot\omega_{\rm GR}}
 {\dot\omega_{\rm CL,bulk}}\right]^{1/5}.
 \label{eq:fr}
\end{equation}
If every other input were fixed, $f_R$ is the radius scale factor required for the classical term to exhaust the nominal excess. Of 26 finite cases, the median is 1.97; one, three, and 14 objects have $f_R\leq1.2$, 1.5, and 2.0, respectively. CH~Ind and SW~CMa have $f_R=1.71$ and 1.97 (Fig.~\ref{fig:fr}). The required changes are much larger than ordinary 10-per-cent radius corrections, but equation~\eqref{eq:fr} is only a screening statistic because $k_2$, rotation, and both component weights evolve with radius. It identifies systems where an incorrect luminosity class or evolutionary state can dominate the claimed excess.

A validation hierarchy should therefore first obtain component masses and radii, then measured or bounded rotations, an evolutionary-state-dependent pair of $k_2$ values at a common age and composition, and finally a covariance-aware timing analysis whose independent units are observing epochs or sectors. As the time span grows, a full nonlinear apsidal timing model should replace a local linear drift \citep{DimoffOrosz2023}. Only the residual after these steps should be mapped to a companion mass--separation relation. LTTE, radial velocities, spectra, and future Gaia epoch astrometry remain essential to distinguish a planet from a brown dwarf, star, or a more complex inclined configuration \citep{Thornton2026}. The screening statistic identifies priorities for this target-specific work; the other candidates still require individual analysis.

\section{Conclusions}

\begin{enumerate}
 \item CH~Ind's target-specific stellar prediction is consistent with the sector-level TESS measurement; no additional prograde apsidal contribution is detected in the nearly coplanar configuration tested here.
 \item SW~CMa's published target-specific theory and observed long-baseline rate already agree, and the bulk search underestimated its classical term by a factor of about 16.
 \item Apsidal-motion surveys can efficiently select CBP targets, but confirmation requires component-level binary parameters and an evolutionary stellar model; the $f_R$ screen provides a lightweight way to prioritize that work.
\end{enumerate}

\section*{Acknowledgements}
This paper includes data collected with the TESS mission, obtained from the MAST data archive at the Space Telescope Science Institute (STScI). Funding for US Institutions for the TESS mission is provided by the NASA Explorer Program. STScI is operated by the Association of Universities for Research in Astronomy, Inc., under NASA contract NAS 5-26555. We acknowledge the use of TESS High Level Science Products produced by the Quick-Look Pipeline at the TESS Science Office at MIT, which are publicly available from MAST. This research has made use of the VizieR catalogue access tool, CDS, Strasbourg, France (DOI: 10.26093/cds/vizier). Computations used NumPy 2.2.6 \citep{Harris2020}, SciPy 1.15.3 \citep{Virtanen2020}, Astropy 6.1.7 \citep{Astropy2022}, Lightkurve 2.6.0 \citep{Lightkurve2018}, and Matplotlib 3.10.7 \citep{Hunter2007}. Generative-AI tools were used to assist with code implementation and language editing. The scientific analysis, interpretation, and conclusions were independently validated by the author, who takes full responsibility for the content of this work. The author declares no conflicts of interest.

\section*{Data availability}
The eclipse timings, sector-level timing nodes, processing variants, stellar-model inputs and outputs, Monte Carlo and robustness scripts, candidate sensitivity table, environment specification, and random seeds underlying this work are available on Zenodo at \mbox{\href{https://doi.org/10.5281/zenodo.21919189}{https://doi.org/10.5281/zenodo.21919189}}. TESS light curves are publicly available from the Mikulski Archive for Space Telescopes (MAST).

\bibliographystyle{mnras}
\bibliography{references}

\end{document}

\maketitle

\section{CH Ind eclipse timing}
\label{sec:timing}

\subsection{Light curves, templates, and individual times}

The preferred light curve uses the \texttt{SAP\_FLUX} column from the Quick-Look Pipeline (QLP; \citealt{Huang2020}) products of the Transiting Exoplanet Survey Satellite (TESS; \citealt{Ricker2015}), with quality-flagged and non-finite samples removed. The processed-flux comparison uses \texttt{KSPSAP\_FLUX} in sectors 1, 27, and 28, and \texttt{DET\_FLUX} in sectors 68, 95, and 104, where \texttt{KSPSAP\_FLUX} is absent. TESS times are stored as BTJD and converted by $\BJD={\rm BTJD}+2457000$. Primary denotes the deeper eclipse. The linear reference ephemerides use $P=5.95257$~d, $T_{\rm pri}=2457000+3155.84289$, and $T_{\rm sec}=2457000+3158.65493$.

Separate empirical templates were built for primary and secondary eclipses. For each predicted centre we extracted $|t-t_{\rm pred}|<0.44$~d. A robust linear polynomial, iteratively clipped at four median absolute deviations (MAD), was fitted to the local continuum at $|t-t_{\rm pred}|>0.31$~d. The normalized eclipse deficit was divided by its local depth, folded about zero, binned every 0.005~d in absolute time, and represented by a shape-preserving cubic interpolant. The template was forced to reach zero deficit at 0.38 and 0.44~d.

Each eclipse centre was then obtained from samples within 0.36~d using
\begin{equation}
 f(t)=c+s(t-\delta)-A\,T(|t-\delta|),
 \label{eq:eclipsefit}
\end{equation}
where $T$ is the appropriate empirical deficit template and the four fitted parameters are centre shift $\delta$, continuum $c$, local slope $s$, and depth $A$. A soft-$L_1$ loss with transition scale twice the typical point uncertainty limits the influence of pulsation residuals and outliers. The centre was restricted to $|\delta|<0.08$~d and solutions with $|\delta|>0.075$~d were rejected. The covariance from the local Jacobian, scaled by the residual variance, gives the formal centre error; one fifth of a cadence is imposed as a lower bound.

For each sector and eclipse type we use the median individual O-C. Its uncertainty is the largest of (i) the quadrature sum of individual errors divided by the number of eclipses, (ii) $1.4826\,\mathrm{MAD}/\sqrt{N}$, and (iii) 0.05~min. This produces six primary and six secondary nodes from the 45 accepted eclipse times. The machine-readable individual and sector tables are included in the data repository described in the Letter.

For the preferred pulsation test, sinusoids at 2.7486 and 8.8527~d$^{-1}$ were fitted outside $|t-t_{\rm eclipse}|<0.34$~d after removing a half-day median trend. Only the sinusoidal terms were subtracted. Repeating the analysis with 10 frequencies and with no pre-whitening tests the sensitivity to this choice; the two-frequency model is not intended to describe the full 46-frequency pulsation spectrum of \citet{Liakos2025}.

\subsection{Differential and joint apsidal models}

To first order in eccentricity for an edge-on binary,
\begin{equation}
 (O-C)_{\rm pri}=-\frac{P}{\pi}e\cos\omega,\qquad
 (O-C)_{\rm sec}=+\frac{P}{\pi}e\cos\omega.
\end{equation}
Hence $D=[(O-C)_{\rm pri}-(O-C)_{\rm sec}]/2=-(P/\pi)e\cos\omega$ and
\begin{equation}
 \dot\omega\,[\mathrm{deg\,cycle^{-1}}]
 =\frac{180}{e\sin\omega}\frac{{\rm d}D}{{\rm d}t}.
 \label{eq:conversion}
\end{equation}
Both $D$ and $t$ are measured in days, so their derivative is dimensionless. We draw $e\sim\mathcal{N}(0.058,0.005^2)$, $\omega\sim\mathcal{N}(222\degr,1\degr{}^2)$, and propagate $e\sin\omega$ in every posterior draw. The measured inclination is $89.41\pm0.01\degr$ \citep{Liakos2025}. Inclination does not enter equation~\eqref{eq:conversion} at first order; the small departure from $90\degr$ and higher-order eccentricity terms are below the present sector-level timing uncertainty. We therefore label this explicitly as a first-order measurement rather than an exact nonlinear apsidal solution.

The six differential nodes are fitted with $D(t)=D_0+b(t-\bar t)$. The node variance is $\sigma_D^2+j^2$, where $j\geq0$ is a common sector jitter with prior $j\sim\mathrm{HalfNormal}(\sigma_j=1~\mathrm{min})$. The numerical marginalization covers $0\leq j\leq4$~min. The intercept and slope have flat priors and are integrated analytically conditional on $j$. The preferred rate also includes a Gaussian method term equal to the standard deviation, $0.178\,\rateunit$, of the four SAP/QLP raw/clean posterior medians.

As a check, all individual times are fitted to
\begin{equation}
 t_{k,s}=T_c+P(k+s/2)+(2s-1)\frac{P}{\pi}e
 \cos[\omega_0+k\dot\omega],
 \label{eq:jointtiming}
\end{equation}
where $s=0$ for primary and 1 for secondary. The fitted parameters are $T_c$, $P$, $e$, $\omega_0$, and $\dot\omega$; angles are converted to radians when the cosine is evaluated. The Gaussian priors on $e$ and $\omega_0$ are those above. Numerical box priors are $3155.74289<T_c<3155.94289$ BTJD, $5.94<P<5.97$~d, $0.02<e<0.12$, $200\degr<\omega_0<245\degr$, and $-20<\dot\omega/(10^{-3}\,\mathrm{deg\,cycle^{-1}})<20$. Uncertainties come from 2500 perturbations in which each sector--eclipse-type group receives a common draw from its sector-node error. This fit gives $1.98^{+0.81}_{-0.72}\,\rateunit$ and is a robustness check, not the reported posterior.

\begin{table*}
\centering
\caption{CH~Ind timing variants. Rates are in $10^{-3}$~deg cycle$^{-1}$ and quote the median and central 68 per cent interval after sector-jitter marginalization.}
\label{tab:timingvariants}
\begin{tabular}{lccccl}
\toprule
Flux and treatment & Rate & $j_{50}$ (min) & $j_{95}$ (min) & Independent nodes & Role\\
\midrule
SAP, raw & $1.85^{+0.98}_{-0.93}$ & 0.20 & 0.79 & 6 sectors & robustness\\
SAP, two frequencies & $1.82^{+0.97}_{-0.93}$ & 0.20 & 0.79 & 6 sectors & preferred\\
\texttt{KSPSAP\_FLUX}/\texttt{DET\_FLUX}, raw & $2.18^{+1.17}_{-1.13}$ & 0.30 & 1.10 & 6 sectors & robustness\\
\texttt{KSPSAP\_FLUX}/\texttt{DET\_FLUX}, two frequencies & $2.09^{+1.18}_{-1.12}$ & 0.31 & 1.11 & 6 sectors & robustness\\
SAP, quadratic continuum & $1.85^{+0.96}_{-0.93}$ & 0.20 & 0.78 & 6 sectors & robustness\\
SAP, ten frequencies & $1.95^{+0.99}_{-0.94}$ & 0.20 & 0.79 & 6 sectors & robustness\\
Joint first-order model & $1.98^{+0.81}_{-0.72}$ & -- & -- & 12 sector--type groups & robustness\\
\bottomrule
\end{tabular}
\end{table*}

\begin{figure*}
 \includegraphics[width=\textwidth]{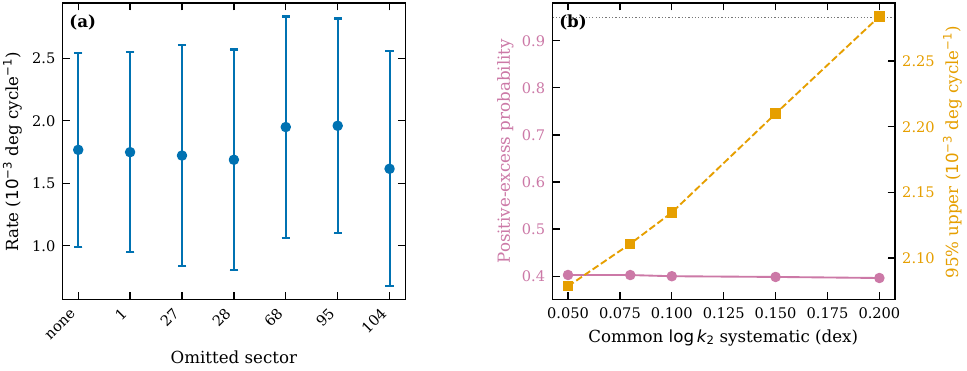}
 \caption{Left: leave-one-sector-out CH~Ind rates. Right: sensitivity to the adopted common $\log k_2$ systematic.}
 \alttext{Alt text: All sector-removal rates overlap, and increasing the stellar-model systematic widens the upper limit without producing a significant positive excess.}
 \label{fig:robustness}
\end{figure*}

Figure~\ref{fig:robustness} repeats the fit after removing each sector. The inferred rate remains consistent across all six omissions, and none gives a significant excess. The seven minima tabulated by \citet{Liakos2025} are not added because they are measurements of the same Sector~68 light curve. The only other catalogued CH~Ind epoch has no primary/secondary classification, uncertainty, or documented time standard, and is not usable as an apsidal datum.

\subsection{The reported CH Ind LTTE}

\citet{Thornton2026} reported an LTTE semi-amplitude of $2.0\pm0.4$~min and an outer period of $10\,056\pm1020$~d. Using $M_1+M_2=3.67\,\mathrm{M}_{\odot}$, the central values give
\begin{align}
 a_{\rm AB}\sin i_3&=cA_{\rm LTTE}=0.240~\mathrm{au},\\
 f(M_3)&=\frac{4\pi^2(a_{\rm AB}\sin i_3)^3}{GP_3^2}
       =1.83\times10^{-5}\,\mathrm{M}_{\odot}.
\end{align}
The corresponding edge-on minimum mass is $M_{3,\min}=0.063\,\mathrm{M}_{\odot}$ ($66.5\,\mathrm{M}_{\rm Jup}$), with a relative outer semimajor axis of 14.1~au. Under the circular, coplanar quadrupole scaling used by \citet{Thornton2026}, this object produces only $1.61\times10^{-6}$~deg cycle$^{-1}$ of inner apsidal motion, about 622 times below their nominal $1.0\times10^{-3}$~deg cycle$^{-1}$ excess. At fixed outer period, even the formal large-$M_3$ limit of the same scaling is $9.46\times10^{-5}$~deg cycle$^{-1}$. Thus the reported LTTE and apsidal excess cannot be attributed to one circular, nearly coplanar companion within that approximation.

This comparison addresses whether one companion can explain both reported signals; it does not test the LTTE detection itself. The proposed period is 27.5~yr, whereas the available TESS baseline samples less than one third of a cycle, and the individual times selected by \citet{Thornton2026} are not public. A new Keplerian LTTE fit would therefore be dominated by the adopted ephemeris and long-term trend. Longer common-mode eclipse timing, systemic radial velocities, or astrometry is needed to test the LTTE orbit.

\section{CH Ind coeval $k_2$ inference}
\label{sec:k2}

We use the MESA tables of \citet{Claret2023}, VizieR catalogue J/A+A/674/A67. For each of $[\mathrm{Fe/H}]=-0.5,0.0,+0.5$, the public 1.8 and $2.0\,\mathrm{M}_{\odot}$ tracks are resampled uniformly in age from 0.30 to 1.60~Gyr and interpolated linearly in mass. The two stars share age and metallicity. Their masses have the independent Gaussian priors in Table~\ref{tab:k2inputs}; the two radii and temperatures have Gaussian likelihoods. Equal prior probability is assigned to the three discrete metallicity grids and age is uniform. Surface gravity and luminosity are not included because they are derived from the same mass, radius, and temperature measurements.

For a model state $\Theta=(\tau,Z,M_1,M_2)$,
\begin{align}
 \ln p(\Theta|\mathbf{y})={}&-{1\over2}\sum_{i=1}^{2}
 \left[\frac{(M_i-M_{i,\rm obs})^2}{\sigma_{M_i}^2}
 +\frac{(R_i-R_{i,\rm obs})^2}{\sigma_{R_i}^2}\right.\nonumber\\
 &\left.\hspace{36mm}+\frac{(T_i-T_{i,\rm obs})^2}{\sigma_{T_i}^2}\right]+C.
 \label{eq:k2like}
\end{align}
Each state supplies separate $k_{2,1}$ and $k_{2,2}$. These are apsidal-motion constants; a fluid Love number defined as twice the apsidal constant must be divided by two before use in the classical formula. The grid has one overshooting prescription, so the adopted common 0.05--0.20-dex shifts are sensitivity tests, not a measured multi-model uncertainty. For each tested $\sigma_{\rm sys}$ we draw one $\Delta\log_{10}k_2\sim\mathcal{N}(0,\sigma_{\rm sys}^{2})$ and add that same shift to both component values in a Monte Carlo realization.

\begin{table}
\centering
\caption{Observed CH~Ind inputs to the coeval fit. Star 1 is the brighter, larger component.}
\label{tab:k2inputs}
\begin{tabular}{lcc}
\toprule
Quantity & Star 1 & Star 2\\
\midrule
$M$ ($\mathrm{M}_{\odot}$) & $1.82\pm0.01$ & $1.85\pm0.02$\\
$R$ ($\mathrm{R}_{\odot}$) & $3.05\pm0.09$ & $2.79\pm0.09$\\
$T_{\rm eff}$ (K) & $6900\pm200$ & $6908\pm75$\\
\bottomrule
\end{tabular}
\end{table}

\begin{table*}
\centering
\caption{Coeval posterior and solar-metallicity best fit. Parentheses in the best-fit columns are residuals in observational standard deviations. The total $\chi^2=8.79$ includes the two mass residuals.}
\label{tab:k2posterior}
\begin{tabular}{lcccc}
\toprule
Quantity & Posterior median & 68 per cent interval & Best star 1 & Best star 2\\
\midrule
Shared age (Gyr) & 1.240 & 1.203--1.267 & \multicolumn{2}{c}{1.244}\\
Mass ($\mathrm{M}_{\odot}$) & -- & -- & 1.808 ($-1.17\sigma$) & 1.801 ($-2.46\sigma$)\\
Radius ($\mathrm{R}_{\odot}$) & -- & -- & 3.081 ($+0.34\sigma$) & 2.854 ($+0.71\sigma$)\\
$T_{\rm eff}$ (K) & -- & -- & 6896 ($-0.02\sigma$) & 6972 ($+0.86\sigma$)\\
$k_2$ & -- & -- & 0.00274 & 0.00244\\
Posterior $k_{2,1}$ & 0.00272 & 0.00251--0.00297 & -- & --\\
Posterior $k_{2,2}$ & 0.00243 & 0.00223--0.00268 & -- & --\\
\bottomrule
\end{tabular}
\end{table*}

The metallicity weights are 0.925, less than $10^{-5}$, and 0.075 for solar, metal-poor, and metal-rich grids. The best solar fit has $\chi^2=8.79$ for three nominal degrees of freedom ($p=0.032$). Most of the tension comes from moving the second mass from $1.85$ to $1.801\,\mathrm{M}_{\odot}$ so that the more massive observed component is not forced to be the larger model component. This is why the inferred constants are not described as precision measurements.

Matching the stars separately gives ages of 1.264 and 1.181~Gyr, $k_{2,1}=0.00232$, $k_{2,2}=0.00238$, and $\dot\omega_{\rm stars}=1.85^{+0.35}_{-0.28}\,\rateunit$ under the same 0.08-dex sensitivity prescription. This also overlaps the observed rate, so the conclusion is not tied to the modest tension in the coeval fit.

The published covariance of the two component masses is unavailable. The baseline therefore uses their quoted errors independently. Holding both masses fixed increases the median stellar rate to $2.91\,\rateunit$; assigning a trial correlation $\rho=+0.5$ gives $1.98\,\rateunit$, compared with $2.05\,\rateunit$ for independent errors. These are sensitivity cases, not alternative covariance measurements. The probability of positive extra precession is 0.17, 0.43, and 0.40 for fixed, correlated, and independent cases, respectively.

\begin{table*}
\centering
\caption{Dependence on the adopted common $\log k_2$ systematic. Rates are in $10^{-3}$~deg cycle$^{-1}$. The final upper limit is the largest non-negative 95 per cent limit across the table.}
\label{tab:k2systematics}
\begin{tabular}{ccccc}
\toprule
Systematic (dex) & Stellar rate (68 per cent) & Extra rate (68 per cent) & $P(\dot\omega_{\rm extra}>0)$ & Non-negative 95 per cent upper limit\\
\midrule
0.05 & $2.051\ (1.774,2.385)$ & $-0.250\ (-1.252,0.787)$ & 0.403 & 2.08\\
0.08 & $2.051\ (1.707,2.485)$ & $-0.258\ (-1.302,0.800)$ & 0.402 & 2.11\\
0.10 & $2.053\ (1.660,2.568)$ & $-0.271\ (-1.346,0.814)$ & 0.400 & 2.13\\
0.15 & $2.051\ (1.535,2.803)$ & $-0.298\ (-1.501,0.853)$ & 0.398 & 2.21\\
0.20 & $2.057\ (1.421,3.083)$ & $-0.331\ (-1.707,0.895)$ & 0.396 & 2.28\\
\bottomrule
\end{tabular}
\end{table*}

\section{Formula and convention audit}

The GR and classical functions return degrees per anomalistic binary cycle. The compact GR coefficient was checked against the SI expression $6\pi G(M_1+M_2)/[ac^2(1-e^2)]$ and agrees to $2.1\times10^{-5}$ fraction, set by rounding of the compact coefficient. Swapping component labels leaves both classical sums unchanged to machine precision. A finite-difference injection of $1.820\rateunit$ into $D=-(P/\pi)e\cos\omega$ is recovered by equation~\eqref{eq:conversion} with a fractional error of $1.8\times10^{-5}$.

The classical calculation uses $n=2\pi/P$, $\Omega_i/n$, and aligned spin and orbital axes. CH~Ind adopts $\Omega_i/n=1.0\pm0.1$ for both stars; this is a sensitivity assumption, not a measured rotation. The anomalistic period differs from the quoted sidereal period by only 2.9~s at the inferred CH~Ind rate, negligible for the budget but retained as a convention distinction. The stellar numbering is fixed throughout as listed in Table~\ref{tab:k2inputs}.

In the published form of equation~(10) of \citet{Thornton2026}, the tidal coefficient contains $M_{3-i}$ rather than the standard mass ratio $M_{3-i}/M_i$ used in equation~(5) of the Letter. The omitted denominator is most naturally read as a typographical error, since the coefficient must be dimensionless. A separate denominator in the rotational term was corrected between the arXiv and journal versions. The component-level bulk inputs and implementation are not available, so it is not possible to determine which tidal expression was used in the numerical calculation. We therefore take the classical rates directly from their Table~1 rather than reconstructing them from the printed equation.

For scale, an equal-component reconstruction using the published GR rates and the target eccentricities gives component masses of 1.150 and $1.362\,\mathrm{M}_{\odot}$ for CH~Ind and SW~CMa. With all other quantities held fixed, the literal printed expression makes the total synchronous classical term 1.133 and 1.334 times the standard result. Thus the published classical values 0.070 and $0.020\,\rateunit$ would become 0.079 and $0.0267\,\rateunit$ if their table used the standard expression but equation~(10) were applied literally; conversely, they would become 0.0618 and $0.0150\,\rateunit$ if the table used the literal expression and the denominator were restored. The table precision and unpublished component inputs do not distinguish these cases. Either correction is much smaller than the factors of 24 and 16 produced by replacing the bulk stellar estimates with measured component dimensions.

\section{SW CMa checks}

\begin{figure}
 \includegraphics[width=\columnwidth]{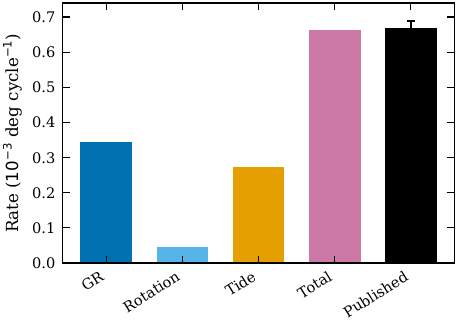}
 \caption{SW~CMa unit test using the effective classical total. The calculation agrees with the published prediction to within one per cent.}
 \alttext{Alt text: Two theoretical total rates overlap.}
 \label{fig:swtest}
\end{figure}

The $\bar{k}_2$ reported for SW~CMa is the coefficient-weighted mean
\begin{equation}
 \bar{k}_2=\frac{\sum_i k_{2,i}(c_i^{\rm rot}+c_i^{\rm tide})}
 {\sum_i(c_i^{\rm rot}+c_i^{\rm tide})}.
\end{equation}
Because the two $k_{2,i}$ values were not published separately, we compute only
$\dot\omega_{\rm CL}=360\bar{k}_2\sum_i(c_i^{\rm rot}+c_i^{\rm tide})$ for the scientific comparison. \citet{Torres2012} obtained the theoretical, coefficient-weighted value $\log\bar{k}_{2,{\rm theo}}=-2.582\pm0.050$ from stellar models, not by inverting the observed apsidal rate. Using its central value and the \citet{Torres2012} binary parameters gives $0.3193\,\rateunit$; adding GR gives $0.6642\,\rateunit$, 0.87 per cent below the $0.670\pm0.020\,\rateunit$ target-specific prediction of \citet{Claret2021} (Fig.~\ref{fig:swtest}). A rotation--tide split obtained by assigning the same effective $\bar{k}_2$ to both stars is useful only for code diagnostics and is not quoted as a physical decomposition.

For the quoted SW~CMa excess posterior we draw $\log\bar{k}_{2,{\rm theo}}\sim\mathcal{N}(-2.581,0.015^2)$ following \citet{Claret2021} and one shared $\Delta\log_{10}k_2\sim\mathcal{N}(0,0.05^2)$, while propagating the measured binary parameters. The 0.05-dex term is a conservative sensitivity allowance, not an empirical calibration.

\begin{figure*}
 \includegraphics[width=\textwidth]{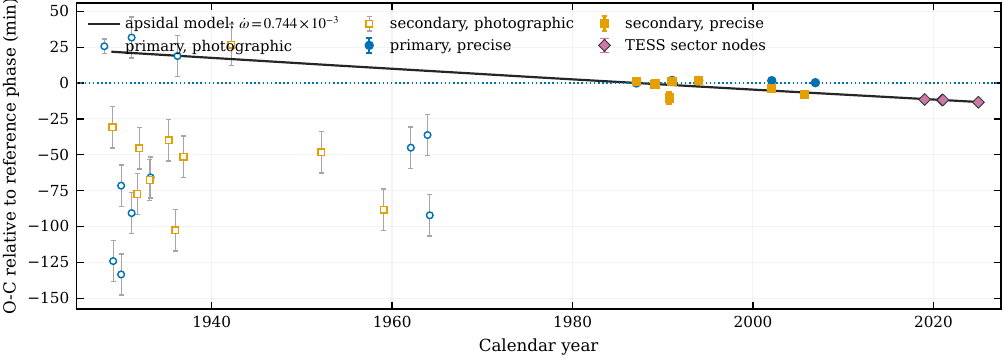}
 \caption{SW~CMa eclipse timings relative to the secondary-eclipse phase at the reference epoch. Open points are low-precision photographic minima; filled points are the precise historical subset, and diamonds are sector-level TESS differences. The preferred rate is the full target-specific solution of \citet{Claret2021}.}
 \alttext{Alt text: Century-long eclipse timings connect to four TESS sector nodes but retain substantial scatter.}
 \label{fig:swhistory}
\end{figure*}

Figure~\ref{fig:swhistory} shows the historical timing cross-check. The early minima are legacy HJD values without a specified underlying time scale. They were not converted to $\BJD$, because an assumed conversion would give them unwarranted precision. We therefore retain the $0.690\pm0.050\,\rateunit$ rate of \citet{Claret2021} for the apsidal budget and use our simplified fit, which has reduced $\chi^2=4.42$, only as a consistency check.

\section{Candidate radius sensitivity}

Table~\ref{tab:fr} gives
$f_R=[(\dot\omega_{\rm obs}-\dot\omega_{\rm GR})/\dot\omega_{\rm CL,bulk}]^{1/5}$
for all 27 candidates, together with the rates used to calculate it. The same
values are supplied in machine-readable form in the data repository. The
diagnostic is undefined for the one object whose tabulated bulk classical term
was rounded to zero.

\begin{table*}
\centering
\caption{Radius-sensitivity diagnostic for the 27 final candidates of \citet{Thornton2026}. Rates are in $10^{-3}$ deg cycle$^{-1}$. The diagnostic is undefined when the tabulated bulk classical term rounds to zero.}
\label{tab:fr}
\begin{tabular}{lrrrr@{\qquad}lrrrr}
\toprule
TIC & $\dot\omega_{\rm obs}$ & $\dot\omega_{\rm GR}$ & $\dot\omega_{\rm CL}$ & $f_R$ & TIC & $\dot\omega_{\rm obs}$ & $\dot\omega_{\rm GR}$ & $\dot\omega_{\rm CL}$ & $f_R$\\
\midrule
123951716 & 3.840 & 0.136 & 0.003 & 4.15 & 146204045 & 7.200 & 0.205 & 0.005 & 4.26\\
SW CMa & 0.850 & 0.253 & 0.020 & 1.97 & 283651681 & 1.100 & 0.240 & 0.030 & 1.96\\
293225466 & 13.300 & 0.036 & 0.000 & -- & 61656788 & 1.790 & 0.420 & 0.200 & 1.47\\
457579488 & 0.480 & 0.182 & 0.010 & 1.97 & CH Ind & 1.300 & 0.290 & 0.070 & 1.71\\
158330804 & 0.610 & 0.190 & 0.032 & 1.67 & 252497283 & 2.700 & 0.300 & 0.100 & 1.89\\
286310830 & 1.120 & 0.220 & 0.020 & 2.14 & 196989952 & 2.570 & 0.420 & 0.400 & 1.40\\
274182408 & 1.900 & 0.260 & 0.060 & 1.94 & 167699456 & 4.000 & 0.110 & 0.003 & 4.19\\
292357653 & 0.507 & 0.150 & 0.020 & 1.78 & 343127696 & 22.200 & 0.310 & 0.200 & 2.56\\
124282654 & 2.900 & 0.172 & 0.008 & 3.21 & 399127035 & 8.800 & 0.260 & 0.100 & 2.43\\
51629874 & 14.700 & 0.230 & 0.040 & 3.25 & 355503224 & 0.800 & 0.260 & 0.040 & 1.68\\
396170777 & 1.300 & 0.250 & 0.050 & 1.84 & 95622298 & 5.900 & 0.320 & 0.100 & 2.24\\
56023695 & 27.300 & 0.390 & 1.200 & 1.86 & 64366964 & 11.500 & 0.490 & 5.900 & 1.13\\
253270207 & 5.100 & 0.340 & 0.100 & 2.17 & 291751499 & 16.200 & 0.300 & 0.110 & 2.70\\
82893635 & 0.760 & 0.130 & 0.010 & 2.29 &  & & & & \\
\bottomrule
\end{tabular}
\end{table*}

\bibliographystyle{mnras}
\bibliography{references}